# Filling-dependent intertwined electronic and atomic orders in the flat-band state of 1*T*-$TaS_2$

Yanyan Geng[1,2,+], Haoyu Dong[1,2,+], Renhong Wang[1,2,+], Jianfeng Guo[1,2,3], Shuo Mi[1,2], Le Lei[1,2,3], Yan Li[3], Li Huang[3], Fei Pang[1,2], Rui Xu[1,2], Weiqiang Yu[1,2], Hong-Jun Gao[3], Wei Ji[1,2,*], Weichang Zhou[4,*], and Zhihai Cheng[1,2,*]

[1]*Key Laboratory of Quantum State Construction and Manipulation (Ministry of Education), Renmin University of China, Beijing, 100872, People's Republic of China*

[2]*Beijing Key Laboratory of Optoelectronic Functional Materials & Micro-nano Devices, Department of Physics, Renmin University of China, Beijing 100872, People's Republic of China*

[3]*Beijing National Laboratory for Condensed Matter Physics, Institute of Physics, Chinese Academy of Sciences, Beijing 100190, People's Republic of China*

[4]*Key Laboratory of Low-dimensional Quantum Structures and Quantum Control of Ministry of Education, School of Physics and Electronics, Institute of Interdisciplinary Studies, Hunan Normal University, Changsha 410081, People's Republic of China*

**Abstract:** The delicate interplay among the complex intra-/inter-layer electron-electron and electron-lattice interactions is the fundamental prerequisite of these exotic quantum states, such as superconductivity, nematic order, and checkerboard charge order. Here we explore the filling-dependent multiple stable intertwined electronic and atomic orders of flat-band state of 1*T*-$TaS_2$ encompassing hole order, phase orders, coexisting left- and right-chiral orders and mixed phase/chiral orders via scanning tunneling microscopy (STM). Combining first-principles calculations, the emergent electronic/atomic orders can be attributed to the weakening of electron-electron correlations and stacking-dependent interlayer interactions. Moreover, achiral intermediate ring-like clusters and nematic charge density wave (CDW) states are successfully realized in intralayer chiral domain wall and interlayer heterochiral stacking regions through chiral overlap configurations. Our study not only deepens the understanding of filling-dependent electronic/atomic orders in flat-band systems, but also offers perspectives for exploring exotic quantum states in correlated electronic systems.

[+]These authors contributed equally: Yanyan Geng, Haoyu Dong, Renhong Wang,

*Email: zhihaicheng@ruc.edu.cn, wchangzhou@hunnu.edu.cn, wji@ruc.edu.cn

The transition metal dichalcogenides (TMDs) deserve particular attention as they exhibit rich electronic states such as charge density wave (CDW) [1-3], superconductivity (SC) [4,5], quantum spin liquid [6], magnetism [7], etc. Most of these electronic states depend on the competition, coexistence or cooperation of electron−electron correlations and electron-lattice interactions and, therefore, are significantly sensitive to variations in the electronic structures [8,9]. Hole or electron doping has been an effective way to tune electronic structures. For example, CDW is very rapidly suppressed in hole Sn-doped $CsV_3Sb_5$ [10], superconductivity occurs in electron/hole-doped twisted graphene moiré superlattice system [11,12], and superconductivity is enhanced in electron K-doped FeSe [13]. Meanwhile, precise carrier doping can also gradually modulate the internal electron-electron correlation, leading to the induction of intriguing quantum states, including checkerboard charge order [14,15], nematic phase [16], a superconducting dome [17], a correlation-driven insulating phase [18,19] and a metallic phase [20,21]. Therefore, a detailed study of the structure and properties of the doping-related electrons is a necessary prerequisite for achieving the controllability of CDW-based exotic quantum states.

Among various TMDs materials, 1*T*-$TaS_2$ has attracted wide attention due to its abundant CDW phase [22-24]. The formation of the Star of David (SoD) in the CDW state creates a half-filled flat band, which is transformed into a Mott insulator or band insulator by electronic correlations or interlayer coupling [25-27]. Meanwhile, because of the close proximity of the various competing interactions in energy, doped electron or hole carriers can modify the flat-band filling factor and effectively modulate the electron correlation effects in CDW to achieve a rich variety of quantum states [28-30]. Prominent examples include the metallization can be realized without breaking the long-range CDW order when surface electron doping is applied [31,32], while hole doping is known to drastically suppress the transition from a nearly commensurate CDW (NCCDW) phase to a commensurate CDW (CCDW) phase [33], and superconductivity coexists with CDW [34]. Furthermore, doping not only can directly affect the interior electron-electron correlations but also induce significantly changes in the interlayer CDW stacking order [22]. Despite extensive previous studies on doping, it is still important to understand the evolution of filling-dependent electronic and atomic orders in the flat-band state based on intralayer and interlayer interactions.

**In this work,** we investigate the filling-dependent intertwined electronic and atomic orders in the flat-band state and internal interlayer/intralayer interactions of 1*T*-$TaS_2$ by STM combined with theoretical calculations. With decreasing flat-band filling factor, a variety of interleaving orders are observed: hole order, phase orders, stacking disorder and intertwined phase/chiral orders in sequence. The emergence of stable fragmentized intralayer phase domain can be attributed to the reduced electron-electron correlations, which is incapable of persisting long-range intralayer orders. The further decreased electron-electron correlations and interlayer interactions contributed to the following appearance of intralayer chiral domains and interlayer heterochiral stacking. At the intralayer chiral domain wall and interlayer heterochiral stacking regions, the intermediate ring-SoD clusters and nematic CDW states are discovered and discussed based on the transient chiral-overlapping interactions. This work provides important insights into the internal interactions and the future realization of exotic quantum states in corrected electronic materials.

## Results and Discussion

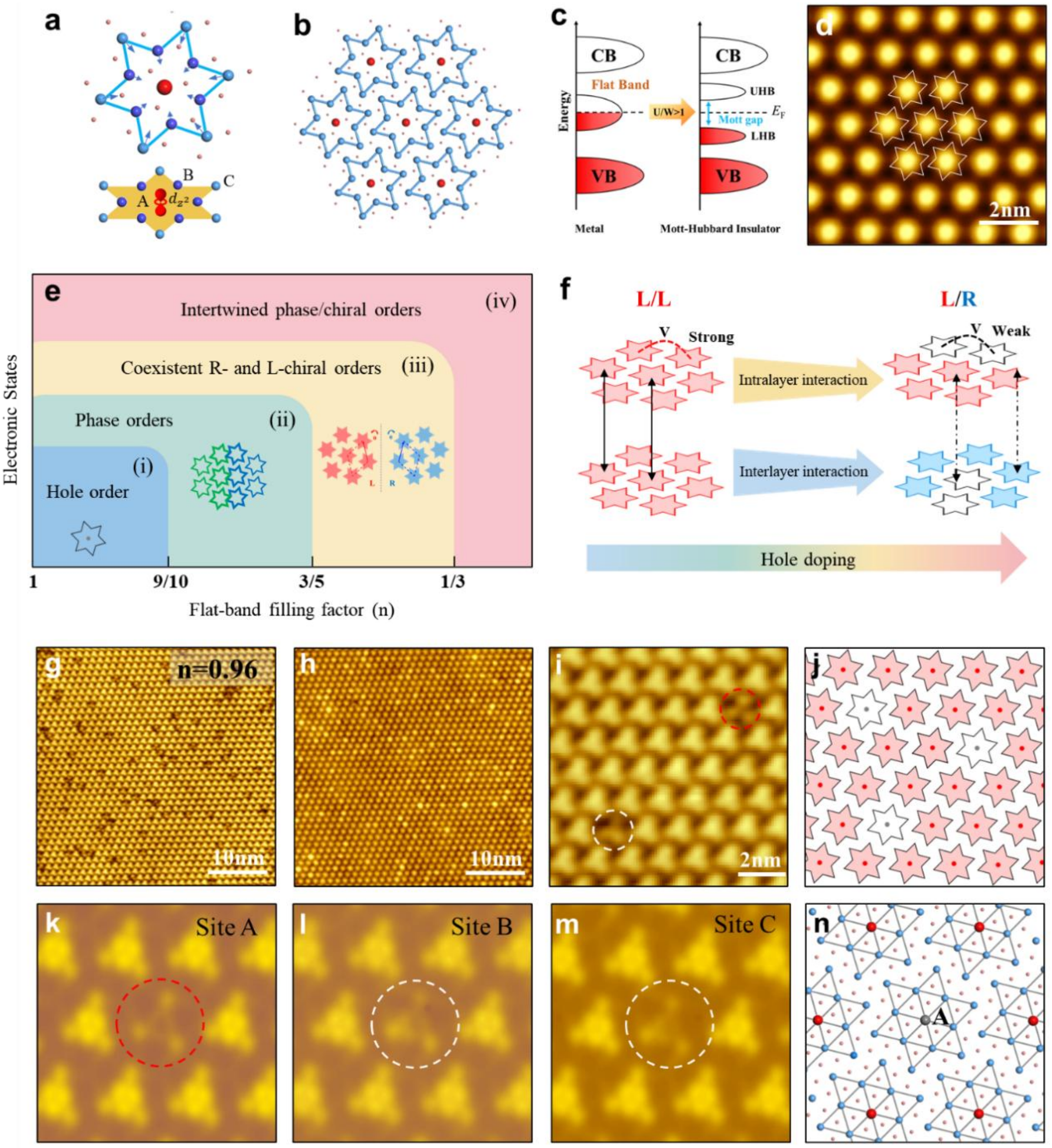

**Figure 1. Structure model of filling-dependent intertwined electronic and atomic orders of 1*T*-$TaS_2$.** (a) Atomic models of the SoD cluster with the localized $d_{z^2}$ orbital (unpaired electron) of central Ta atom. (b) Schematic of the SoD clusters with the $\sqrt{13} \times \sqrt{13}$ periodicity. (c) Schematic band structures of correlated CDW state, in which the half-filled flat band of the localized $d_{z^2}$ orbital is split into UHB and LHB bands via electron correlations. (d) STM image of commensurate CDW state with the overlaid SoD models. (e) Schematic diagram of electronic orders and flat-band filling factors of 1*T*-$TaS_2$. Each SoD is fully central electron-filled, corresponding to a flat-band filling factor n = 1 for 1*T*-$TaS_2$. While each SoD is fully central electron-empty, corresponding to a flat-band filling factor n = 0 for 1*T*-$TaS_2$. (f) Schematic of the delicate interplay of the complex intralayer and interlayer interactions with the hole doping. (g-i) Large-scale occupied state (g), unoccupied state (h), high-resolution (i) STM images of hole order at the flat-band filling factor n=~0.96. (j) Structure model of electron-filled and -empty SoDs. The central

electron-filled and -empty SoDs are marked by the central red and grey dots, respectively. (k-m) Simulated STM images of the three inequivalent substitution sites: site A(k), site B(l), and site C(m). (n) Structure model of the inequivalent substitution site A. Scanning parameter: (g, i) $V$=-0.4 V, $I$=-100 pA; (h) $V$=0.4 V, $I$=100 pA.

The ground CDW state of 1$T$-$TaS_2$ is characterized by the commensurate $\sqrt{13}\times\sqrt{13}$ superlattice of SoD clusters, as depicted in Fig. 1a-b. The SoD cluster is made of 12 inward-distorted outer Ta atoms and one center Ta atom, in which the twelve $5d_{z^2}$ electrons of the outer Ta atoms pair and form six occupied CDW valence bands. The remaining one unpaired electron of SoD is maximally localized at the central Ta atom to form the half-filled flat band. A relatively strong intralayer electron-electron correlations then leads to the split upper Hubbard band (UHB) and lower Hubbard band (LHB), forming a correlated Mott insulator without considering the interlayer interaction (Fig. 1c). Recently, the important role of the interlayer stacking order has been emphasized, and the ground CCDW state of 1$T$-$TaS_2$ can be described as a trivial band insulator/conventional Mott insulator on the termination of $T_A$/$T_c$ stacking at low temperatures [25-27]. Figure 1d presents the STM image of the ground CCDW state with the overlaid star models of SoDs.

Hole doping can effectively modify the flat-band filling factor, providing an exemplary platform to fine-tune the electronic/atomic structure and realize a wide range of intriguing electronic states. Figure 1e illustrates the schematic diagram of electronic orders and flat-band filling factors, which shows the evolution of a series of electronic orders: (i) hole order, (ii) phase orders, (iii) coexisting left- and right-chiral orders, and (iv) intertwined phase/chiral orders. Notably, the phase and chiral domains created by hole doping can reach the millimeter scale in a stable state, in contrast to the various metastable domain induced by pressure, ultrafast optical, and electrical pulses [35-39]. These emergent electronic states can be phenomenologically attributed to the competition and/or cooperation of stacking-dependent interlayer interactions and intralayer interactions of 1$T$-$TaS_2$ (Fig. 1f). In 1$T$-$TaS_2$, with the increase of hole doping, the reduction of intralayer electrons contributes to the gradually weakening of electron-electron correlation, resulting in the inability of the CCDW to maintain long-range ordering and the emergence of hole and phase/chiral domain walls (DWs). Meanwhile, hole doping is also accompanied by the gradual diminishing of interlayer SoD interactions (T) (Fig. S1) [22], which leads to the emergence of random stacking order and interlayer heterochiral stacking.

Compared to the pristine 1*T*-$TaS_2$, the slight reduction of flat-band filling electrons results in the appearance of hole-SoDs within the CCDW superlattice, showing darker (brighter) contrast than the normal SoDs at the occupied (unoccupied) state, as shown in Figs. 1g and 1h. These hole-SoDs are clearly resolved as the darker three-petal flower shapes in the high-resolution STM image of Fig. 1i, and further schematically illustrated in Fig. 1j. To determine the origin of the three-petal flower SoD, first-principles calculations and d*I*/d*V* spectra are performed. It can be seen that the typical d*I*/d*V* spectra of bright triangular SoDs and dark three-petal flower SoDs show significant differences [Fig. S2]. The bright SoDs display an electron-hole symmetric gap, while the dark SoDs exhibit an electron-hole asymmetric single pronounced peak that onsets sharply above the Fermi energy. According to the first-principles simulation, the three-petal flower SoDs can be well reproduced by selecting the central (site A) or peripheral Ta (site B or C) as the Ti substitution sites as the shown in Fig. 1k-1n and Fig. S3. Since the central (site A) has the lowest doping energy (Supplementary Table I), we suggest that the three-petal flower SoDs originate from the substitution of Ti atoms for the central Ta atoms, leaving the SoDs without central Mott electrons.

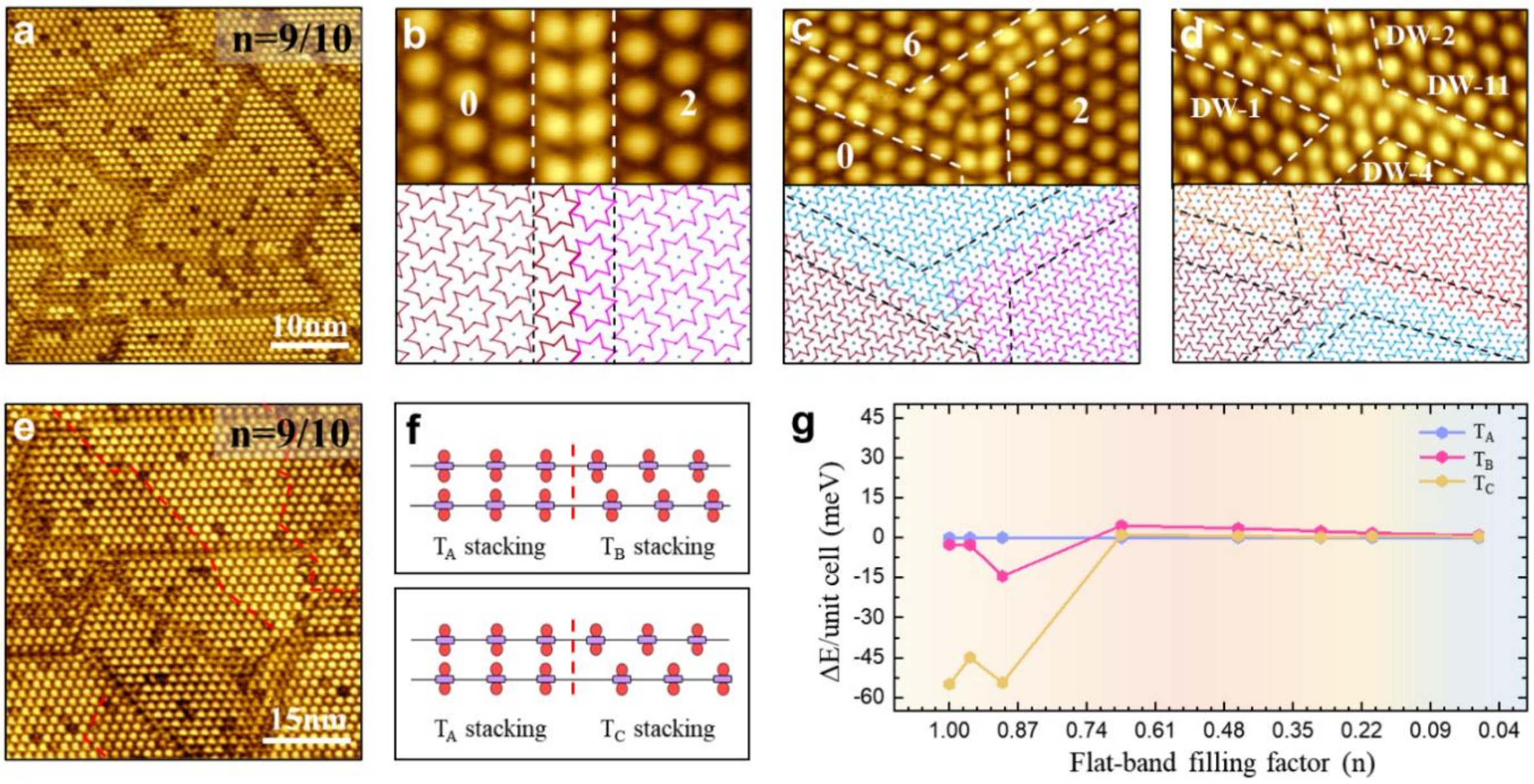

**Figure 2. Emergence of phase domains at the flat-band filling factor n=~9/10 of 1*T*-$TaS_2$.** (a) Large-scale STM images of the emergent phase domains of 1*T*-$TaS_2$. (b) STM images of the typical CDW

phase DWs (top) and their schematic structural configurations (bottom). The constituent SoDs of phase DWs are highlighted by the dashed lines separating the two neighboring phase domains. (c,d) STM images of two typical CDW vortexes and their schematic structural configurations. (e) STM images for sublayer phase DWs. The red lines denote the phase DWs of the sublayer. (f) Arrangement of the localized states of the central Ta atoms for the three representative CDW stacking configurations: $T_A$, $T_B$, and $T_C$. (g) First-principles calculations of the $T_A$, $T_B$, and $T_C$ stacking configurations as a function of flat-band filling factor in 1*T*-$TaS_2$. Scanning parameter: (a, e) *V*=-0.4 V, *I*=-100 pA; (b-d) *V*=0.4 V, *I*=100 pA.

The obvious effect of the reduction of the flat-band filling factor is to create domains of diverse shapes and sizes separated by sharp DWs (Fig. 2a). Most DWs are formed straightly along the CCDW unit vector with a single-phase shift between two adjacent CCDW domains as shown in Fig. 2b and Figs. S4-S5. In addition, the DWs also construct the complex DW junctions including the X, Y, and so on (Figs. 2c and 2d). It is worth noting that the mosaic-like CDW states are the stable ground state in the thermal equilibrium, which is existent in each layer and different from the metastable (by the external energy excitations) mosaic state of surface layers in pristine 1*T*-$TaS_2$ in recent reports (Fig. S6) [37].

Interestingly, in addition to the DWs in the surface layer, there are extra but weaker domain walls within a relatively large single-domain (indicated by the red dashed lines in Fig. 2e), which is assigned as the sublayer DWs [37, 42]. These subdomains correspond to different interlayer CDW stacking configurations (Fig. 2f), resulting in different apparent contrasts in the surface-sensitive STM images. The calculation results (Fig. 2g and Fig. S7) show that the flat-band filling factor has a dramatic effect on the stability of interlayer CDW stacking configurations. In the case of a flat-band filling factor n = 1, $T_C$ stacking dominates. As the filling factor decreases, the energy of $T_C$ stacking gradually increases (Supplementary Table II). At the flat band filling factor n =~3/5, the energies of the various stacking configurations are similar and the stacking order is random. It means that doping reduces the stacking-dependent interlayer interactions, leading to various local interlayer stacking configurations, which increases the complexity of these mosaic-like CDW states.

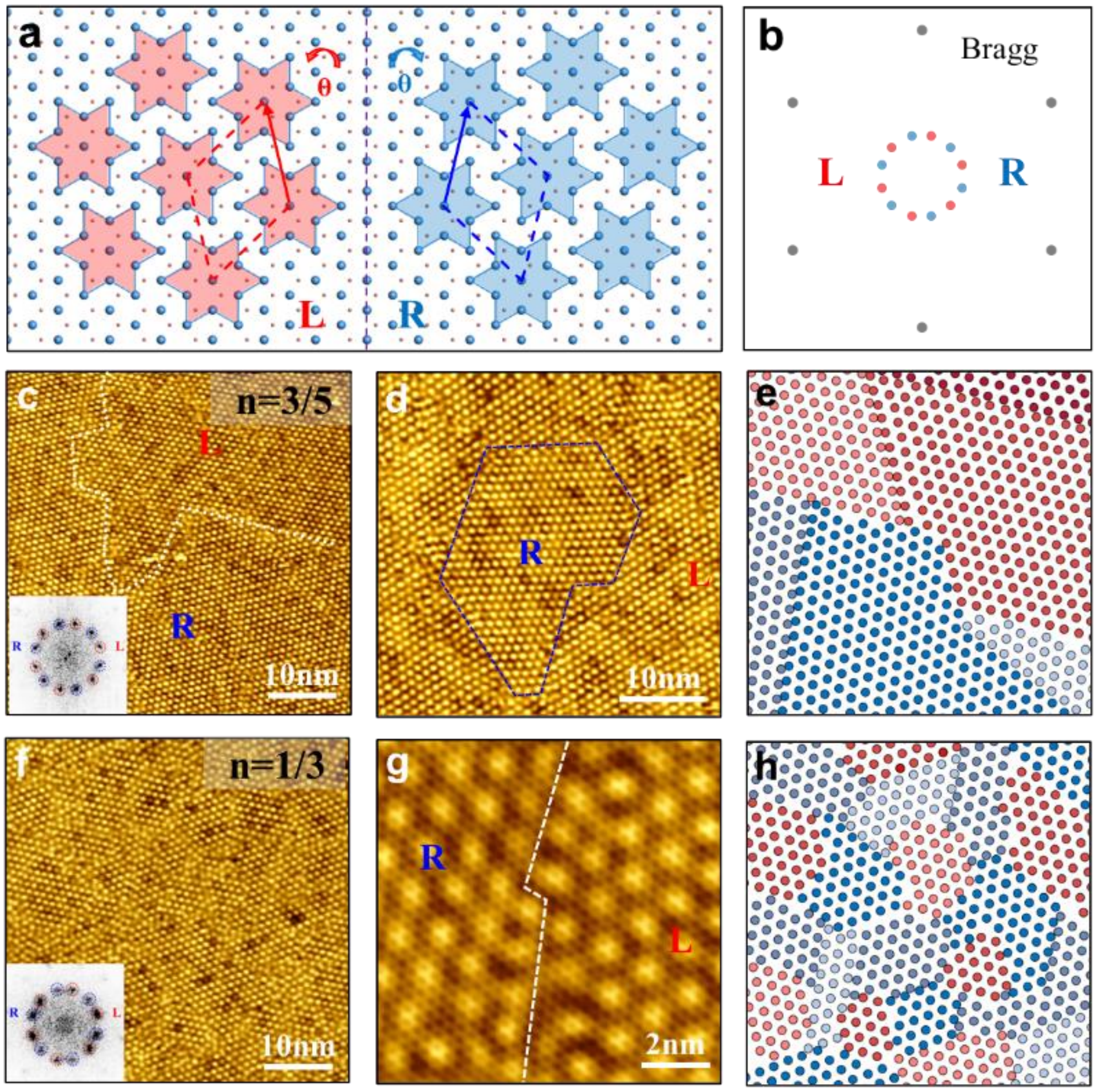

**Figure 3. Filling-dependent intertwined phase and chiral orders of 1*T*-TaS₂.** (a) Schematic of L- and R- chiral CDW domains. Unit cells of L- and R-chiral domains are shaded in red and blue, respectively. (b) Schematic FFT pattern of L- and R-chiral CDW domains with the atomic Bragg spots. (c,d) Large-scale STM images of the emergent chiral domain wall at the flat-band filling factor n=3/5 of 1*T*-$TaS_2$. The chiral DW is along the [110] directions of R-chiral domain and marked by blue dashed lines in (d). (e) The structural model of chiral domains. The emergent large chiral-arranged CDW domain is made of many small homochiral phase domains. (f) Large-scale STM images of intertwined chiral and phase domains at the flat-band filling factor n=~1/3 of 1*T*-$TaS_2$. (g) Atomic-resolution STM image of typical chiral DWs. The chiral DW is marked by white dashed lines. (h) The structural model of mixed chiral/phase domains. The fragmentized CDW state is made of randomly distributed tiny phase and chiral domain. Scanning parameter: *V*=0.4 V, *I*=100 pA.

In 1*T*-$TaS_2$, SoD can also be arranged into different chiral domains: L-chiral and R-chiral, as shown schematically in Figs. 3a and 3b. The SoD superlattices are rotated clockwise from the Ta atomic lattice by -13.9° and +13.9° in the L- and R-chiral-arranged CDW domains, respectively. In general, the lateral size of the $\sqrt{13}\times\sqrt{13}$ domain of a SoD superlattice in $1T$-$TaS_2$ is much larger than normal scan range of STM measurements of about a few hundred nanometers [40], and their interlayer stackings are almost exclusively preferred to the homochiral stacking even in different phase stacking configurations. Only very few works about the chiral CDW state were just reported recently with the focus on their global chiral-switching via the circularly polarized light or electrical pulse in the nanothick flakes [41,42].

At the flat band filling factor n =~3/5, the large chiral CDW domains emergent within the mosaic-like CDW states, as shown by the marked chiral DWs to separate the R- and L-chiral domains in Figs. 3c-d and the corresponding FFT pattern. The large chiral domains are made of many small homochiral phase domains, as schematically illustrated in Fig. 3e. It is also noted that different from the DWs at low-doping-level, the DWs could not be readily resolved based on their specific local contrast. At the flat-band filling factor n=~1/3, the fragmentized CDW states are observed, made of randomly distributed tiny phase and chiral domains, as shown in Figs. 3f and 3h. The short phase and chiral DWs are disordered and could not be clearly defined. The atomic-/SoD-resolution characterizations of these fragmented DWs were further performed, as depicted in Fig. 3g, indicating the perfect continuous atomic lattice across the DWs without any visible doping-induced atomic defects and/or dislocations.

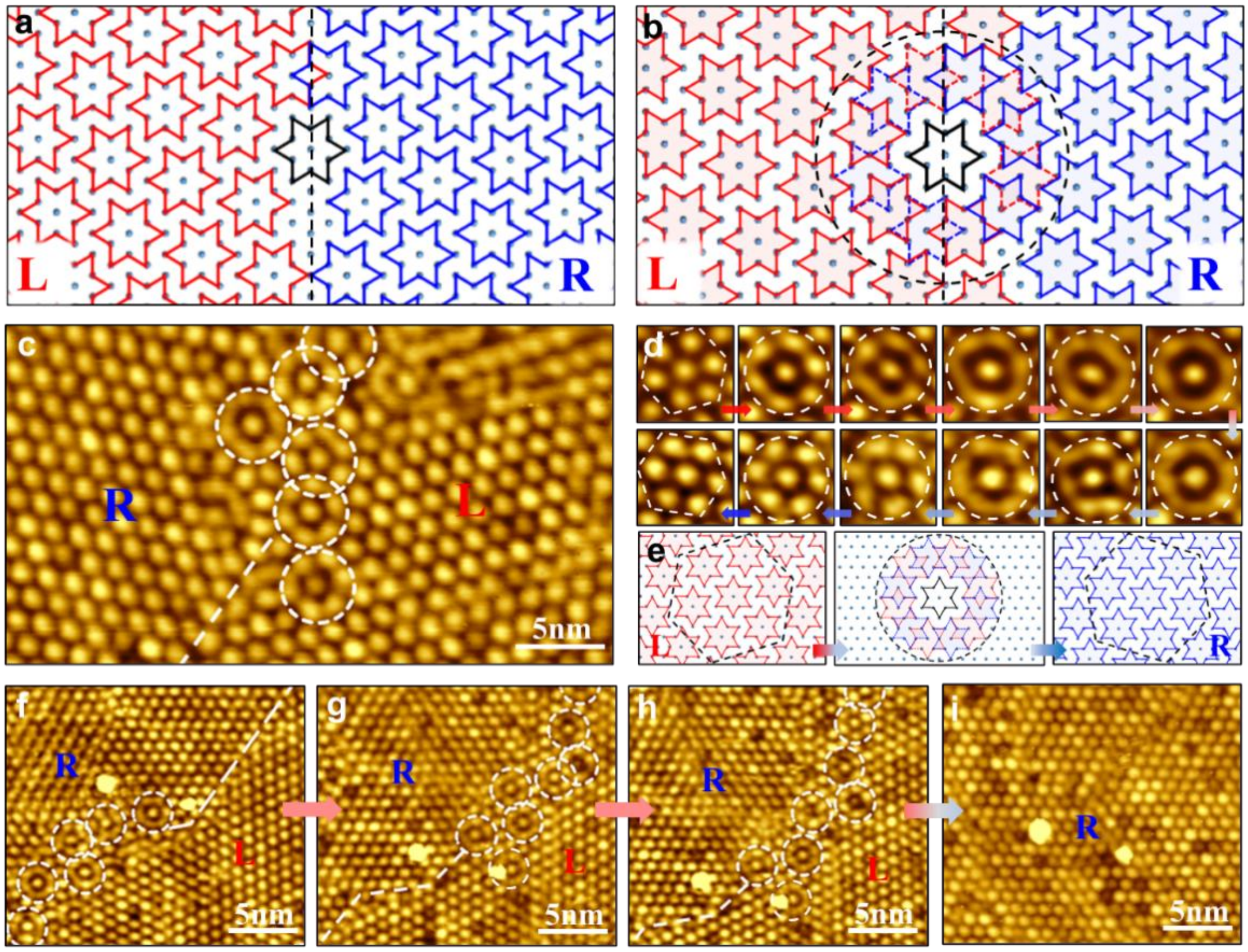

**Figure 4. Emergence of intralayer ring-SoD clusters on the chiral domain wall of 1*T*-$TaS_2$.** (a) Atomic structural model of chiral DW along the [110] direction. This kind of chiral DW is mostly observed in STM images, which should be relatively stable considering the complex intra-SoD interactions around the chiral domain walls. (b) Chiral SoD-overlapping model of the mirror-symmetric DW with the central achiral SoD (black star). (c) STM images of the emergent intermediate ring-SoD clusters around the chiral DWs of 1*T*-$TaS_2$. (d,e) High-resolution STM images and structural models of a series of dynamic intermediate ring-SoD clusters configurations from L-chiral SoD to R-chiral SoD, phenomenally indicating the structural evolution of SoDs and the process of chiral-switching transition. (f-i) A sequential *in-situ* STM images showing dynamic chiral-switching of ring-SoD clusters at the chiral DWs. The ring-SoD clusters are marked with the dashed circles. Scanning parameter: $V$= 0.4 V, $I$= 100 pA.

The mirror-symmetric chiral DWs along the [110] and [211] directions of the atomic lattice are illustrated in Fig. 4a and Fig. S9. The chiral DWs along the [110] direction is mostly observed in STM images, as shown in Fig. S10, which should be relatively stable considering the complex inter-SoD interactions around the chiral DWs. Particularly, such chiral DWs easily form the intermediate ring-SoD clusters with a unique central achiral SoD as an anchor point as indicated by the black dashed circles in Fig. 4b. By variable-temperature STM, the STM image containing the ring-SoD cluster features are successfully attained at 50 K (Fig. 4c). In contrast to single chiral domains, such ring-SoD clusters are rarely observed and are found only in larger chiral domains with flat-band filling factor n = ~3/5. The ring-SoD clusters (white dashed circles) typically have a lateral size of ~5 × 5 nm and are surrounded by chiral DWs. Interestingly, these ring-SoD clusters act as thermally excited dynamic structures that can move along the chiral DW and act as intermediate electronic states of the SoD during chiral transitions (Fig. S11). The evolution of chiral transitions via intermediate ring-SoD clusters is further visualized in Fig. 4d, which is schematically illustrated by the atomic model in Fig. 4e and Fig. S12.

Next, we concentrate on the evolution of the metastable intermediate ring-SoD clusters. The sequential *in-situ* STM characterizations (identified by the bright spots) provide a detailed insight into the dynamics of the chiral changes in CDW superlattices (Figs. 4f-4i). It can be seen that the two chiral domains change over time, with the R-chiral domain expanding and the L-chiral domain gradually shrinking until it eventually all transforms into a structural domain with a single R chirality (Fig. 4i). Moreover, the atomic arrangement of the CDW superlattice remains continuous and intact after the change of chirality. In 1*T*-$TaS_2$, electron-electron correlation and e-*ph* interaction are key factors in the formation of SoD cluster. As the flat-band filling factor decreases, the number of correlated electrons reduces, leading to a weakening of electron-electron correlation. Meanwhile, the ring-SoD clusters serve as the intermediate states during chiral transitions, causing the redistribution of atoms and the renormalization of the phonon frequency. Therefore, we suggest that the appearance of the ring-SoD structure can be phenomenologically attributed to the electron-electron correlation and e-*ph* coupling.

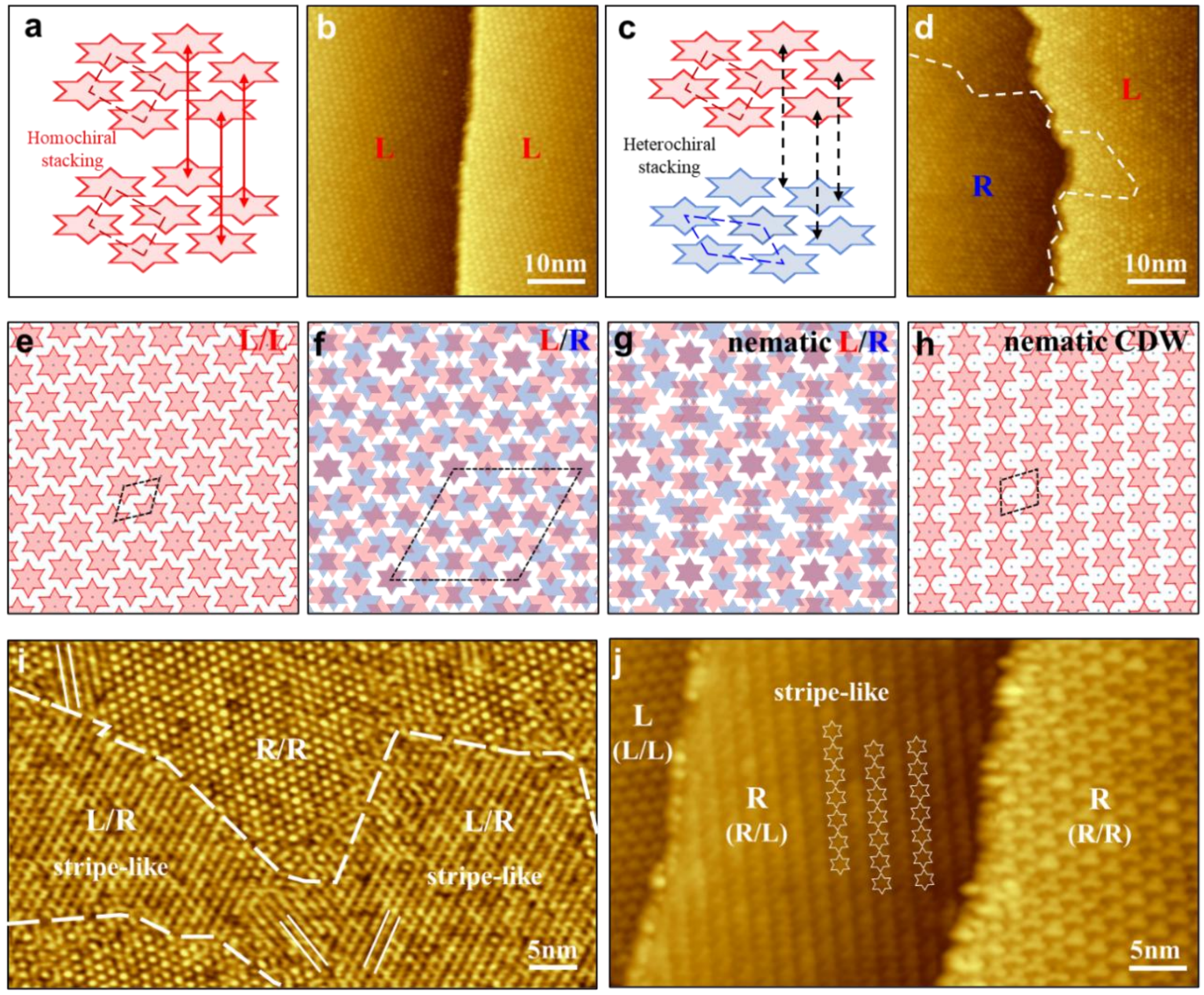

**Figure 5. Interlayer heterochiral CDW stacking and the emergent nematic CDW state of 1*T*-$TaS_2$.** (a,b) Schematics and STM images of step edge in homochiral CDW stacking configurations. (c,d) Schematics and STM images of step edge in heterochiral CDW stacking configurations. (e,f) Illustration of CDW superstructures for homochiral L/L and heterochiral L/R stacking. Unit cells of the CDW superstructures are outlined in black. (g) Proposed directional relaxed heterochiral stacking configurations for the observed stripe-like pattern of nematic L/R stacking. (h) Structural model of the stripe-like nematic CDW state. (i) Large-scale STM image of the chiral domain with the emergent stripe-like nematic CDW state due to heterochiral stacking configuration. (j) STM image of two step edges with respective homochiral/heterochiral stacking configurations and the accompanying stripe-like feature of nematic CDW state. Scanning parameters: (b,d,j) $V$=-0.4 V, $I$=-100 pA; (i) $V$=0.4 V, $I$=100 pA.

In 1*T*-$TaS_2$, the manipulation of interlayer stacking allows for the construction of customizable quantum systems exhibiting exotic physics. The models of the homochiral and heterochiral CDW units are depicted in Figs. 5a and 5c. The STM experiments can easily identify the CDW chirality of both layers by measuring the angle between the close-packed direction of SOD and the step edges, which directly visualizes the two different stacking structures of homochiral L/L and heterochiral L/R CDW stacking order as shown in Figs. 5b and 5d. At near full filling of the flat band, no obvious chiral domains are observed in both the upper and lower layers, showing homochiral stacking (Fig. 5b), which is consistent with pristine 1*T*-$TaS_2$ [43,44]. At half-filling of the flat band, the reduction of electron-electron correlation contributes to the appearance of chiral domains, meanwhile the weakening of interlayer interactions is accompanied by the simultaneous emergence of two different chiral domains in both the upper and the lower layers, resulting in heterochiral stacking (Fig. 5d).

In contrast to the homochiral L/L stacking configuration (Fig. 5e), a 27.8° rotational mismatch between the L- and R-layer CDWs in the heterochiral L/R stacking configuration contributes to a commensurate CDW moiré superstructure (Fig. 5f). In the CDW moiré unit cell (black diamond-shaped box in Fig. 5f), the relaxed atomic structure transforms from a ($\sqrt{13} \times \sqrt{13}$) R13.9° triangular superlattice to a larger 13 × 13 super-superlattice. This superstructure CDW state is energetically very unfavorable and susceptible to spontaneous symmetry breaking, inducing a redistribution of atoms into a nematic structure. We propose the possible formation of the intermediate nematic L/R structure (Fig. 5g) and the final nematic CDW configuration after structural relaxation of the superstructure generated by CDW chiral coexistence (Fig. 5h).

Real-space investigations of the nematic CDW state at heterochiral L/R stacking interfaces are presented in Fig. 5i and Fig. S13. By statistically analyzing the areas occupied by homochiral L/L (R/R) stacking structures and the areas occupied by heterochiral L/R (R/L) stacking configurations, we can roughly determine the ratio of homochirality to heterochiral to be approximately 5:5. One-dimensional stripe-like features with different orientations are observed in the heterochiral L/R stacking region, and the stripes are all along the [211] direction of the underlying atomic lattice. This can be directly evidenced by the observed CDW states around the homochiral and heterochiral stacking edges in Fig. 5j. Note that we can rule out the multiple tip effect because it would have caused discontinuity in the surrounding CDW unicells while the unicells are smoothly connected in the above observation. The detailed mechanism for the formation of the nematic CDW states needs to be further investigated in conjunction with theoretical calculations, which is beyond the scope of this paper.

## Conclusion

$1T$-$TaS_2$ exhibits abundant physical properties and electronic structure, providing an ideal platform to investigate the CDW order and strong electron-electron correlations. In the CCDW state, the central Mott electrons of SoDs form a flat-band structure at the $E_F$. This phenomenon is attributed to the interplay between electron-electron correlations and electron-lattice interactions. The $1T$-$TaS_2$ in the flat-band state demonstrates diverse quantum states, including superconductivity, CDW, and insulating states. The modulation of these quantum states is crucial for advancing our understanding of correlated electronic systems. Hole doping, as a particularly effective means of regulation, not only adjusts the flat-band filling factor in $1T$-$TaS_2$ by altering the electron density, but also induces changes in electronic interactions. This dual effect suggests a promising avenue for research aimed at realizing quantum states.

We directly visualize the filling-dependent stable multiple intertwined orders in $1T$-$TaS_2$, involving hole order, phase domains, chiral domains, and mixed phase/chiral domains by STM (Fig. S14). The evolution of intralayer electron correlations and interlayer CDW stacking order with flat-band filling factor is further investigated in conjunction with DFT calculations. As the flat-band filling factor approaches 1, the system exhibits a long-range ordered CCDW structure with some hole orders. As the flat band filling factor decreases, phase and chiral domains of different shapes and sizes appear. As the flat-band filling factor further decreases, interlayer CDW stacking becomes progressively disordered, contributing to the weakening of interlayer interactions and the emergence of interlayer heterochiral stacking. The interplay of complex electron-electron correlations, electron-lattice interactions, and interlayer interactions collectively induces a redistribution of atoms into a ring-SoD and stripe-SoD ordered structures, accompanied by a change in symmetry. This modulation of the atomic structure further feeds back into the electronic structure, complicating the interactions between electronic and atomic orders. The study paves a way to realize exotic quantum states via the accurate tuning of interior interactions and flat-band filling factors in correlated materials. This approach to manipulating interactions and electronic degrees of freedom will inspire further research focus on electronic structures, properties, and applications in the field of correlated electron materials.

## Methods

**Single crystal growth of Ti-doped 1*T*-$TaS_2$.**

The high-quality 1*T*-$Ta_{1-x}Ti_xS_2$ single-crystals were grown by chemical vapor transport (CVT) method. Ta (99.99%, Aladdin), $TaCl_5$ (99.99%, Aladdin), S (99.99%, Aladdin), and Ti (99.99%, Aladdin) powders with a nominal molar ratio of 1:0.02:2.05:x (x =~ 0.3%, 0.8%, 2.5%, 4.2%, 6.3%) were mixed thoroughly and then loaded into quartz tubes (inner diameter/outer diameter/length: 14/16/200 mm). Large 1*T*-$Ta_{1-x}Ti_xS_2$ crystals with size up to 10 mm were collected for further characterization and measurement. X-ray diffraction (XRD) and X-ray energy dispersive spectroscopy (EDS) were employed to determine the crystal structure, morphology, and composition of as-prepared samples. The Ti-doping contents of all samples agree with their nominal molar ratio of source materials, demonstrating the controllable synthesis. In order to visualize the filling factor of the flat bands more visually for different doping concentrations, we approximate the filling factor n as the fraction.

**Scanning tunneling microscopy (STM).**

High-quality Ti-doped 1*T*-$TaS_2$ crystals were cleaved at room temperature in ultrahigh vacuum at a base pressure of $2\times10^{-10}$ Torr, and directly transferred to the cryogen-free variable-temperature STM system (PanScan Freedom, RHK). Chemically etched W tips were used for STM measurement in constant-current mode. The STM tips were calibrated on a clean Ag(111) surface. Gwyddion was used for STM data analysis.

**Density functional theory calculations and simulations.**

Our DFT calculations were performed using the generalized gradient approximation for the exchange-correlation potential, the projector augmented wave method and a plane-wave basis set as implemented in the Vienna ab-initio simulation package (VASP). Dispersion correction was implemented using the DFT-D3 method, with the PBE functional for the exchange potential. The effective U value of the on-site Coulomb interaction of the Ta *d* orbitals is 2.3 eV. All our simulation cells contain a 20 Å vacuum to prevent interlayer coupling between different layers. The plane-wave cutoff was set to 450 eV. A k-mesh of 5×5×1 was adopted to sample the first Brillouin zone of the cell. Convergence is reached if the consecutive energy difference is within 10−5 eV for electronic iterations and the forces are less than 0.01 eV Å.

## Supporting Information

The supplementary materials and figures show details of STM, STS, and theoretical calculation results including interlayer and interlayer interactions, phase domain walls, chiral domain walls, ring-SoD and stripe-SoD structures, etc.

## Acknowledgments

This project was supported by the National Key R&D Program of China (MOST) (Grant No. 2023YFA1406500), the National Natural Science Foundation of China (NSFC) (No. 92477128, 92477205, 12374200, 11604063, 11974422, 12104504), the Strategic Priority Research Program (Chinese Academy of Sciences, CAS) (No. XDB30000000), and the Fundamental Research Funds for the Central Universities and the Research Funds of Renmin University of China (No. 21XNLG27). Y.Y. Geng was supported by the Outstanding Innovative Talents Cultivation Funded Programs 2023 of Renmin University of China. This paper is an outcome of "Study of Exotic Fractional Magnetization Plateau Phase Transitions and States in Low-dimensional Frustrated Quantum Systems" (RUC24QSDL039), funded by the "Qiushi Academic-Dongliang" Talent Cultivation Project at Renmin University of China in 2024.

## Author contributions

W.J., H.G., W.Z. and Z.C. conceived the research project. Y.G., H.D., L.L. and Z.C. performed the STM experiments and analysis of STM data. J.G., S.M., Y.L., L.H., F.P., R.X. and W. Y. helped in the experiments. Y.G., H.D., W.Z. and Z.C. wrote the manuscript with inputs from all authors.

## Competing Interests

The authors declare no competing financial interests.

## Data Availability

The authors declare that the data supporting the findings of this study are available within the article and its Supplementary Information.